\documentclass[]{aa}
\usepackage{txfonts}
\usepackage{natbib}
\usepackage{graphicx}
\usepackage{psfig}
\usepackage[a4paper]{hyperref}
\begin{document}
\def\teff{$\mathrm T_{eff }$ }
\def\logg {log\,g}
\def\lambo{$\lambda$ Boo }
\def\vsini {$v\,\sin i$ }
\def\kms {$\mathrm{km\, s^{-1}}$ }
\headnote{Research Note}
\title{  
C,N,O  in \lambo stars and in composite spectra 
}
   \subtitle{}
\author{R. \,Faraggiana \inst{1}
\and P. Bonifacio \inst{2}
}
  \offprints{R. Faraggiana}

\institute{
Dipartimento di Astronomia, Universit\`a degli Studi di Trieste,
Via G.B.Tiepolo 11, I-34131 Trieste, Italy \\
%email: faraggiana@ts.astro.it
\and
Istituto Nazionale di Astrofisica --
Osservatorio Astronomico di Trieste,
Via G.B.Tiepolo 11, I-34131 Trieste, Italy 
}

\mail{faraggiana@ts.astro.it}
\authorrunning{R. Faraggiana \& P. Bonifacio}
\titlerunning{C,N,O  in \lambo stars and in composite spectra}

\date{Received ... / Accepted ...}
\abstract{ 
The selective abundance of C, N, O (almost solar) with respect to
that of other elements (underabundant) in \lambo stars has been 
interpreted as a characteristic peculiarity of these objects, when considered
as single stars. We show here that  a similar selective abundance is 
predicted from the composite 
spectra resulting from two unresolved stars in the same temperature range 
as the \lambo stars.

}
\maketitle{}

\section{Introduction}

Weak metal lines were discovered by Morgan et al. (1943) in the \lambo 
spectrum; this star became the prototype of a new class of Ap stars when 
similar characteristics were detected in other A-type stars with 
small space velocity. 

The almost solar abundance of oxygen in the \lambo
stars has been known for more than 40 years (Baschek \& Searle, 1969); however the
solar C,N,O, and S abundances did not enter in the definition of this class
until quite recently.
The approximately solar abundances of C, N, O, and S 
are discussed by Venn \& Lambert (1990) for  3 \lambo stars.
The interpretation, proposed by these authors of the 
dichotomy between the solar abundances of C, N, O, and S 
and those of the other elements
(selective accretion 
of gas, but not dust from circumstellar or interstellar medium)
is still the most widely accepted for these stars when considered as 
single peculiar objects. 

On the basis of this paper by Venn \& Lambert, Paunzen et al. (1997) 
introduced the 
definition of \lambo  class as " Pop I hydrogen burning A-type stars, which
are, except of (sic) C, N, O and S,  metal poor". 
However several stars classified 
\lambo in this paper do not belong to the  A  spectral type; for example, 
the following  stars have \teff 
lower than 7000 K: HD 81290, HD 83041, HD 84948, HD 106223, HD 142994. 
One cause of concern in this definition is that the optical lines
used to derive the abundances of C, N, O are of
high excitation and show a non-monotonic
behaviour with \teff.

In 1999 Faraggiana \& Bonifacio suggested that the \lambo phenomenon could be 
produced by the  
combination of an unresolved binary composed of two stars with not very different
L and \teff. In the subsequent series of papers we demonstrated that this is 
indeed the case for a significant percentage of stars classified as \lambo.
In this paper we investigate theoretically the 
properties of composite spectra for
the optical lines of C,N, O.

\section{CNO line intensity variation with \teff}

We restrict the analysis to C, N, O since the abundance of S is very
uncertain in \lambo stars; Venn \& Lambert (1990) did not measure any SI 
line in  \lambo, and the scatter of the S abundance is 0.5 dex  
between the 4 measured lines  in 29 Cyg and  0.7 dex between the 2 lines 
measured in $\pi^1$ Ori.

Synthetic spectra were computed for solar
abundances of all elements by using Kurucz (1993) model atmospheres 
and SYNTHE code; the atomic line data are taken from the modified Kurucz line
lists available at the CDS and 
http://wwwuser.oat.ts.astro.it/castelli/stars.html 
(see also Castelli \& Kurucz 2004).
All the spectra have been computed by adopting \vsini=50 \kms, which is 
a lower limit for the  \lambo stars (see Table 4 in Gerbaldi et al. 2003).
The selected spectral regions are  those that include the C, N, O  lines 
selected by Venn \& Lambert (1990), and the computations were performed 
over a wavelength
range  broad enough to also include  nearby lines of other "underabundant"
elements. We added the OI 777.4 nm triplet region because this feature is  
often used for oxygen abundance determination. 

The variations of EWs with spectral 
type or \teff were studied by Faraggiana et al. (1988) for oxygen and by 
Takeda \& Takada-Hidai (1995) for nitrogen. 
In the first study the behaviour of the OI 777.4 nm triplet was studied 
and the result of the measured EWs vs. spectral type plotted in Figs. 2 and 3
of that paper. The maximum intensity of this pattern is reached for late
A-type stars of the main sequence. The high sensitivity to the log g parameter
does not alter the bell-shaped behaviour, which is enhanced at lower gravities
when the NLTE effect is included (Baschek et al. 1977).
In the second paper the NI 870.3 and 862.9 nm are  analyzed in A-F supergiants.
The theoretical EWs vs. \teff are plotted in their Fig. 3;
the  bell-shaped variation, similar to that of OI triplet is evident.
For both elements the line intensity reaches the maximum at \teff of 
about 8000 K and decreases for higher and lower temperatures.
A similar study was not made  for CI 
broad blends, as the
carbon abundance was always derived by synthetic spectra analysis.

\begin{figure}
   \centering
   \includegraphics[clip=true]{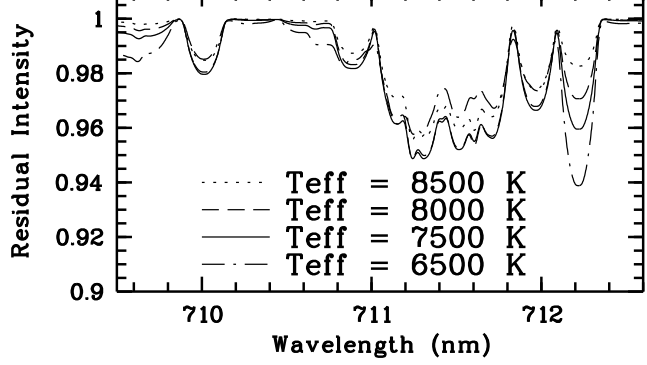}
      \caption{
Variations with \teff of the CI feature 710-712 nm
and that of NiI 712.219 nm. }
         \label{cteff}
   \end{figure}

\begin{figure}
   \centering
   \includegraphics[clip=true]{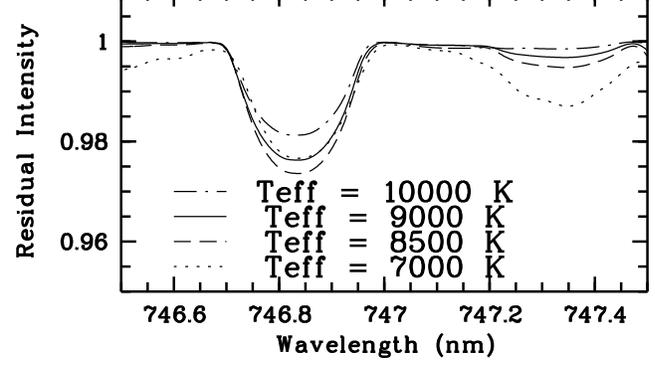}
      \caption{
Variations with \teff of the NI 746.831 nm (blend with SI 746.859)
and the FeI blend 747.275-747.356 nm. }
         \label{nteff}
   \end{figure}

\begin{figure}
   \centering
   \includegraphics[clip=true]{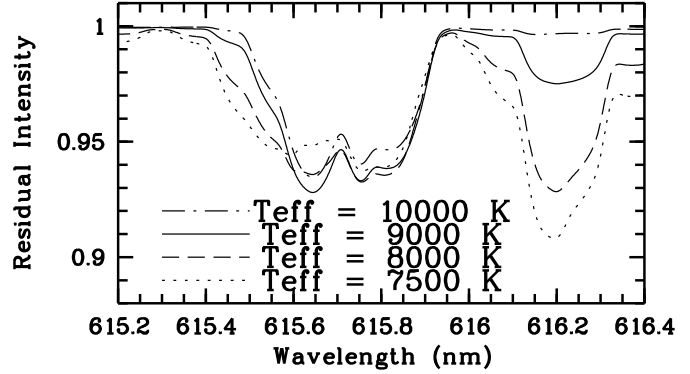}
      \caption{
Variations with \teff of the OI feature 615-616 nm and that of CaI 
616.217 nm. }
         \label{oteff}
   \end{figure}

\begin{figure}
   \centering
   \includegraphics[clip=true]{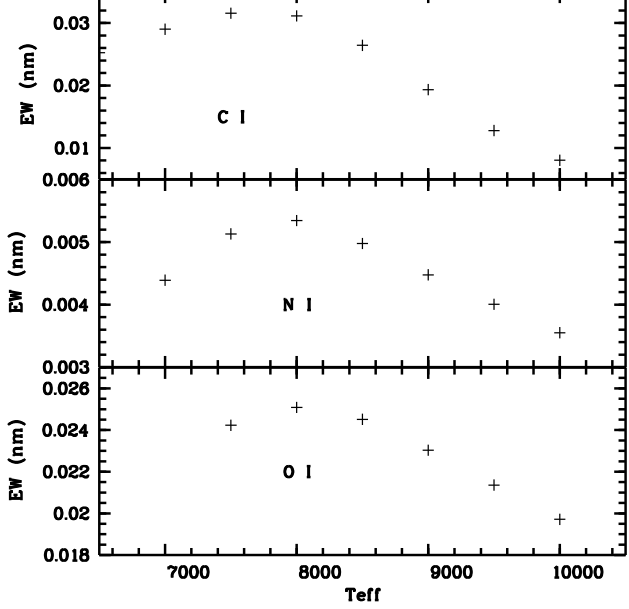}
      \caption{
Variations with \teff of the equivalent width
of the  CI feature at 710-712 nm (top panel), 
  the NI 746.831 nm line (middle panel), and
the OI feature at 615-616  nm. }
         \label{pl_ew}
   \end{figure}

In order to show the behaviour of C, N, O, we  
computed synthetic spectra in the range of \teff between 7000
and 9000 K with 500 K steps and the results are plotted in Figs.
\ref{cteff} to \ref{pl_ew}.
The small variations of C, N, O lines compared to those
of other elements are evident in Figs. 1 to 3.

This result is the consequence of the fact that
the excitation potentials of the lower level of
the neutral C, N,  O lines are rather high
(about three times higher than those of other lines).
Therefore, in going from \teff = 8000 K to 
7500 K, the number densities of the neutral species
{\em increases} because of the shift in the ionization balance,
at the same time the population of the highly excited
levels {\em decreases}; the two effects partially cancel
out so that the line strength is roughly the same.

\section{CNO in composite spectra}

CNO line intensities are almost constant in the \teff range 
7000-9000 K, where most of the \lambo stars lie, and 
are equal for two \teff that are symmetric with respect to the maximum.  
If the composite spectrum is interpreted as due to a single object, the derived
parameters are a weighted average of those of the components, and most 
of the metal lines appear too weak  exception for C, N, O lines,
as observed in \lambo stars.
In order to better illustrate this fact the following simulation was made.

\section{Colours of a binary star}

Given two stars of known \teff and log g, let us
suppose we know the Str\"omgren colours of the two
stars denoted as $A$ and $B$:
$(b-y)_A, m1_A, c1_A$ and
$(b-y)_B, m1_B, c1_B$.
Furthermore suppose we know
the flux ratios of the two stars in the four Str\"omgren bands:
$p = Fu_A/Fu_B$;
$q = Fv_A/Fv_B$;
$r = Fb_A/Fb_B$;
$s = Fy_A/Fy_B$.
Let us then compute the Str\"omgren colours of a  binary system
made up of stars $A$ and $B$.
The quantities referring to the binary system will have no
subscripts, while the ones referring to the single stars
will have subscripts $A$ and $B$, respectively.

By definition 
$\displaystyle b-y = -2.5 \log[(Fb/Fy)/(Fb/Fy)_{st} $,
where subscript $st$ refers to the zero colour standard star
(typically Vega).
Now
$\displaystyle Fb/Fy = (Fb_B/Fy_B)\times{(r+1)\over (s+1)}$;
hence,
$$\displaystyle b-y = (b-y)_B -2.5\log[(r+1)/(s+1)] $$
The definition of m1 is a bit more complex:
$$\displaystyle m1 = (v-b)-(b-y) = -2.5log[(FvFy/Fb^2)/(FvFy/Fb^2)_{st}]$$
It can be shown that
$$ \displaystyle m1 = m1_B -2.5 \log[(q+1)(s+1)/(r+1)^2] $$
Finally, since 
$c1 = (u-v)-(v-b)$,  it can be shown that
$$\displaystyle c1 = c1_B -2.5 \log[(p+1)(r+1)/(q+1)^2]$$

We shall now work out an example which applies the
above formulae. From the isochrone of Girardi et al. (2002) 
with 	Z = 0.01900, age = 10 Myr,
we may read that a star 
of mass 1.5 M\sun~ has \teff = 7185 K, log g = 4.29, and
a star of mass 1.95 M\sun~ has \teff = 8997 K, log g = 4.31.
To compute the colours of a binary system made by
the above two stars, we refer to the 1.5M\sun~ star as $A$
and to the 1.95M\sun~ star as $B$. 
At our request L. Girardi computed the 
Str\"omgren magnitudes for this isochrone,
which allowed us to compute 
the ratios ($p=0.363,q=0.316,r=0.258,s=0.307$)

The Str\"omgren colours of star $B$ may be taken from the
grid by Castelli \& Kurucz (2004) with solar
metallicity and microturbulent velocity of 2 \kms. 
Just reading off this table for
the model with \teff = 9000 K,  log g = 4.5,
one finds $(b-y)_B = 0.035$, $m1_B = 0.211$, and
$c1_B = 0.929$.
For the binary system made of stars A and B we  obtain 
$ b-y = 0.076,  m1 = 0.120, c1 = 0.925$.
The well-known code {\tt TEFFLOGG} of 
Moon \& Dworetsky (1985) also requires knowledge
of the $\beta$ magnitude to determine the
atmospheric parameters of the star; however
since for many stars  the measure of $\beta$ is not available,
the common practice is to estimate $\beta$
using the code {\tt UVBYBETA} of Moon (1985), which also
estimates the reddening. Using the above values
we obtain a very small reddening (E$(b-y) = 0.002$)
and $\beta=2.871$; and by
feeding this estimate and dereddened colours to the
code {\tt TEFFLOGG}, we obtain
\teff = 8330 K, log g = 4.21.

In the V band the luminosity ratio is LA/LB=0.307,     
which implies
LA/Ltot=0.235  and LB/Ltot=0.765.

   \begin{figure}
   \centering
   \includegraphics[clip=true]{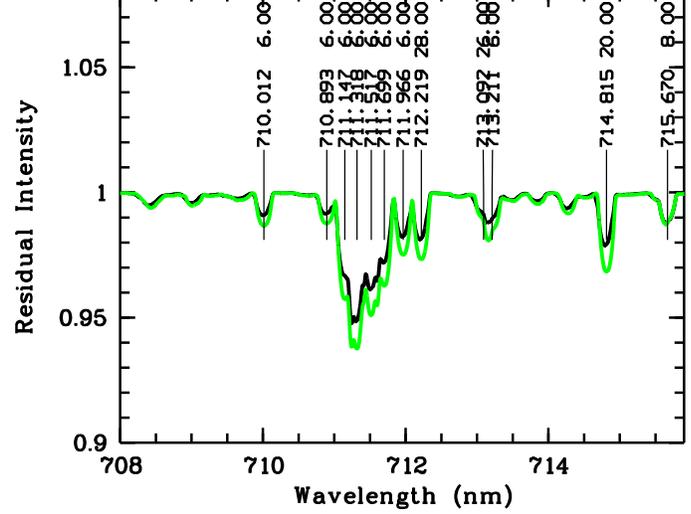}
      \caption{
CI 710.893, 711.147, 711.318, 711.517, 711.699,  and  711.966 nm
in a composite spectrum simulating a \lambo star (black line) and in the single 
one (grey line) based on the average parameters of this binary. Note that 
the very weak (so no-measurable in \lambo stars) CaI 713.843 nm 
arising from a high EP level shows the same behaviour as CI lines.
}
         \label{carbon}
   \end{figure}

   \begin{figure}
   \centering
   \includegraphics[clip=true]{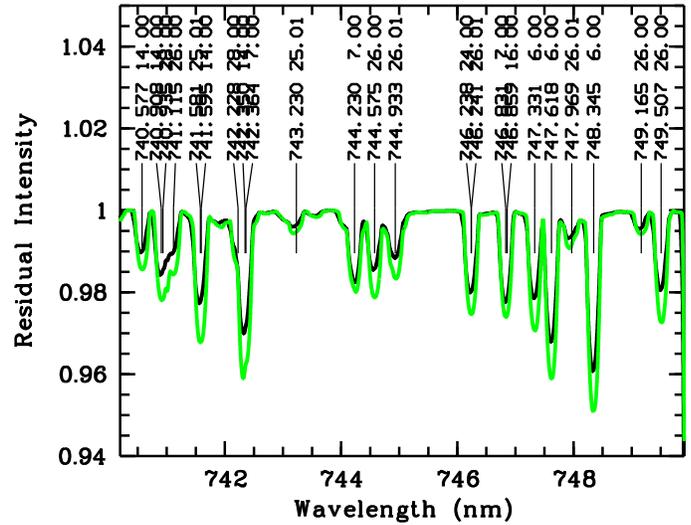}
      \caption{
The NI 744.230  and 746.831 nm in the same couple of 
computed spectra as those in Fig. \ref{carbon}.}
         \label{nitrogen}
   \end{figure}

\begin{figure}
\centering
\includegraphics[clip=true]{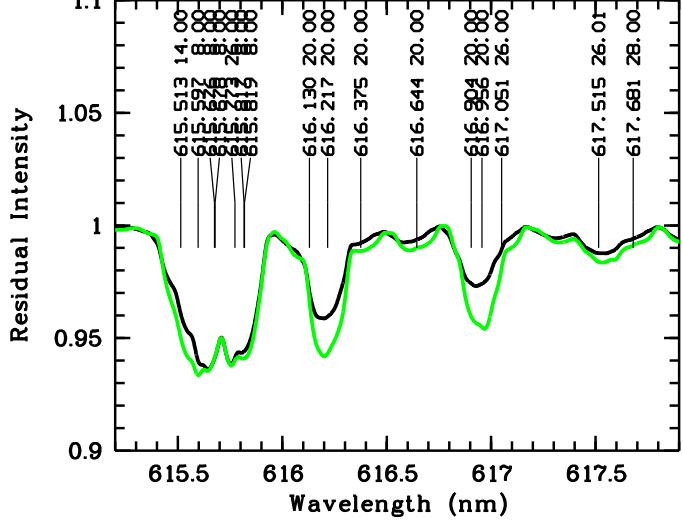}
\caption{
The OI blend 615.597 to   615.819 nm in the same computed 
spectra as those plotted in Fig. \ref{carbon}.
}

\label{oxygen}
\end{figure}

   \begin{figure}
   \centering
   \includegraphics[clip=true]{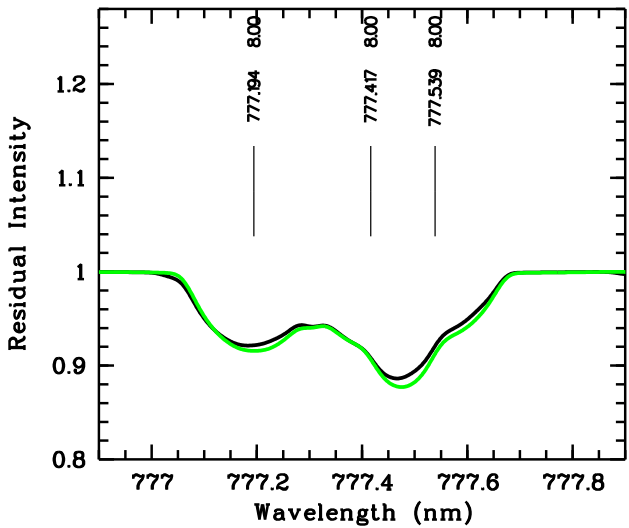}
      \caption{
The OI triplet 777.4 in the same computed spectra as those plotted in Fig. \ref{carbon}}.
         \label{otriplet}.
   \end{figure}

\section {Simulated composite spectrum}
In order to clarify the behaviour of the lines of 
these elements in single and composite spectra
in the \teff domain covered by 
\lambo candidates we made some computed spectra simulations.
We compare the line intensity of C, N, O  lines to those of other elements
in a single spectrum and also in a composite
spectrum, that, according to our binary 
hypothesis, may roughly represent that of a \lambo star.
The chosen \teff and log g of the single spectrum are those derived in the 
previous section, namely 8330 K and 4.21, while  those used to produce the 
composite one are \teff=8997 K, log g =4.31 and \teff=7185 K, log g =4.29. 
 
The combined spectrum that simulates a predicted 
composite spectrum of a binary system was computed by combining the 
two theoretical spectra,  imposing  the luminosity 
ratios LA/Ltot=0.235 and LB/Ltot=0.765 
and a relative shift of 20 \kms 
between the two components. 

The results shown in Figs. \ref{carbon} to \ref{otriplet},
clearly show that lines of elements different from C, N, O 
are stronger in the spectrum computed with the parameters derived from
the average colour indices than in the spectrum obtained by combining the
fluxes of the two single sources. 
 
In contrast, the CNO 
lines have practically the same intensity in both spectra.

\section{Conclusion}
We can conclude that an unresolved binary, 
if erroneously interpreted as a single object,
produces average colours intermediate between those of 
the binary components, while  the corresponding computed spectrum
shows the same kind of abundance peculiarities as a \lambo 
spectrum. 
In particular, the apparently solar abundances of C, N, O  
in \lambo stars cannot be considered a selective criterion to distinguish 
single peculiar stars belonging to the \lambo class.
We have thus shown that the effect of a combined
spectrum is not necessarily to uniformly weaken the lines 
of {\em all elements } by the same  amount.
This implies that the deduced abundance pattern
of a star with a composite spectrum, when interpreted as
a single object, will be altered when compared to the 
real one.

The interpretation of the C, N,  O abundances higher than that of 
Fe-peak elements remains ambiguous: it can be explained either as the
effect of 
the "\lambo phenomenon" in a single star or as due to a binary system
that shows a composite
spectrum but has been analysed as single object.

Therefore the argument that a C,N,  O pattern different from
that of other elements is sufficient to classify a star as
\lambo and to exclude that it is a binary  with a composite spectrum
appears unjustified.  
This argument cannot be used to reject
our interpretation (see  Acke \& Waelkens, 2004)  that many  \lambo stars 
are unresolved binaries with composite spectra.

\begin{acknowledgements}
We are grateful to L. Girardi for providing
Str\"omgren magnitudes for one of his isochrones.
We are also grateful to the language editor J. Adams, who
greatly improved our presentation.
\end{acknowledgements}

\end{document}